# Pressure-induced Superconductivity and Topological Quantum Phase Transitions in the Topological Semimetal ZrTe$_2$

*Shihao Zhu[1#], Juefei Wu[1#], Peng Zhu[2,3,4#], Cuiying Pei[1], Qi Wang[1,5], Donghan Jia[6], Xinyu Wang[6], Yi Zhao[1], Lingling Gao[1], Changhua Li[1], Weizheng Cao[1], Mingxin Zhang[1], Lili Zhang[7], Mingtao Li[6],Huiyang Gou[6], Wenge Yang[6], Jian Sun[8], Yulin Chen[1,5,9], Zhiwei Wang[2,3,4*], Yugui Yao[2,3], Yanpeng Qi[1,5,10]*

[#] These authors contributed to this work equally.

1. School of Physical Science and Technology, ShanghaiTech University, Shanghai 201210, China
2. Centre for Quantum Physics, Key Laboratory of Advanced Optoelectronic Quantum Architecture and Measurement (MOE), School of Physics, Beijing Institute of Technology, Beijing 100081, China
3. Beijing Key Lab of Nanophotonics and Ultrafine Optoelectronic Systems, Beijing Institute of Technology, Beijing 100081, China
4. Material Science Center, Yangtze Delta Region Academy of Beijing Institute of Technology, Jiaxing, 314011, China
5. ShanghaiTech Laboratory for Topological Physics, ShanghaiTech University, Shanghai 201210, China
6. Center for High Pressure Science and Technology Advanced Research, Shanghai, 201203, China
7. Shanghai Synchrotron Radiation Facility, Shanghai Advanced Research Institute, Chinese Academy of Sciences, Shanghai 201203, China
8. National Laboratory of Solid State Microstructures, School of Physics and Collaborative Innovation Center of Advanced Microstructures, Nanjing University, Nanjing 210093, China
9. Department of Physics, Clarendon Laboratory, University of Oxford, Parks Road, Oxford OX1 3PU, UK
10. Shanghai Key Laboratory of High-resolution Electron Microscopy, ShanghaiTech University, Shanghai 201210, China

* Correspondence should be addressed to Y.Q. (qiyp@shanghaitech.edu.cn) or

Z.W. (zhiweiwang@bit.edu.cn)

**Topological transition metal dichalcogenides (TMDCs) have attracted much attention due to its potential applications in spintronics and quantum computations. In this work, we systematically investigate the structural and electronic properties of topological TMDCs candidate ZrTe$_2$ under high pressure. A pressure-induced Lifshitz transition is evidenced by the change of charge carrier type as well as the Fermi surface. Superconductivity was observed at around 8.3 GPa without structural phase transition. A typical dome-shape phase diagram is obtained with the maximum $T_c$ of 5.6 K for ZrTe$_2$. Furthermore, our theoretical calculations suggest the presence of multiple pressure-induced topological quantum phase transitions, which coexists with emergence of superconductivity. The results demonstrate that ZrTe$_2$ with nontrivial topology of electronic states display new ground states upon compression.**

1. Introduction

In the past decades, the layered transition metal dichalcogenides (TMDCs: $M$X$_2$, $M$ = Mo, W, Ta, Zr, Hf, etc., and X = S, Se, or Te) have attracted tremendous attention owing to their rich physics and potential device applications[1-8]. The diversity of electronic properties of TMDCs includes the charge density wave (CDW)[9-11], the magnetism[12-14], and the superconductivity (SC)[15, 16]. Recent studies have shown that TMDCs exhibit nontrivial topology[17-20], making the study of these materials even more intriguing. In particular, superconductivity was observed successfully in TMDCs, either in stoichiometric compounds at ambient or under high pressure, or by doping/intercalation individual layers[21-24]. Therefore, TMDC family provides an exotic platform to study the relation between topologically non-trivial state and superconductivity and even exploration of topological superconductivity (TSC)[25-28].

Among TMDCs, ZrTe$_2$ has been relatively little investigated; however, it is predicted to possess non-trivial band topology recently. Although theoretical calculations indicated that ZrTe$_2$ is a topological crystalline insulator protected by crystalline symmetry[29-31], however, angle-resolved photoemission spectroscopy (ARPES) studies have revealed that ZrTe$_2$ is a topological semimetal with approximately equal electron and hole carrier densities[32, 33]. Tian *et al.* performed nuclear magnetic resonance (NMR) experiments and supported ZrTe$_2$ as a quasi-2D Dirac semimetal with a nodal line between $\Gamma$ and $A$[34]. Negative magnetoresistivity has been observed in both thin films, single crystals and nanoplates prepared by mechanical exfoliation[35-37], further indicating its topological semimetal features. More interestingly, bulk superconductivity was observed by intercalating Cu or Ni in the van der Waals gap, indicating a possible candidate for TSC[22, 38].

The application of pressure can effectively tune the crystal structures and the corresponding electronic states in a valid and systematic fashion, and its related studies on $M$X$_2$, indeed, have given rise to many novel physical phenomena[23, 39-44]. To date, the high-pressure properties of ZrTe$_2$ have not been well explored. Here, we systematically explore the structure and electronic properties of topological TMDC ZrTe$_2$ single crystal under high pressure. Room-temperature synchrotron x-ray diffraction and Raman scattering measurements reveal the stability of the hexagonal CdI$_2$-type structure up to 49.3 GPa. We demonstrate a pressure-induced Lifshitz transition revealed by the sign change of the charge carrier type and the Fermi surface. A superconducting transition is observed in ZrTe$_2$ at around 8.3 GPa and $T_c$ reaches the maximum of 5.6 K around 19.4 GPa. Through the first-principles calculations, we find that the application of pressure alters the electronic properties and leads to multiple topological quantum phase transitions in ZrTe$_2$.

2. **Results and discussion**

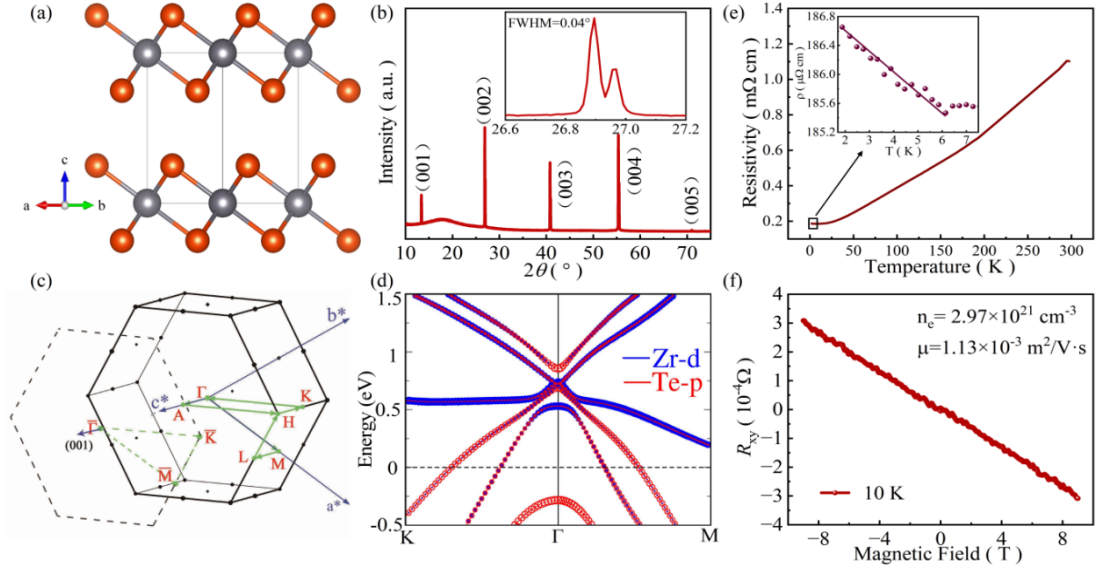

**FIG. 1**. (a) The crystal structure of $ZrTe_2$ (Zr: gray; Te: orange). (b) X-ray diffraction peaks from the ab plane of $ZrTe_2$ single crystal. The inset shows the details of the (002) reflection. (c) The bulk Brillouin zone and its projections onto the conventional cell (001) surface. (d) Calculated band structure of $ZrTe_2$ with spin-orbit coupling (SOC). (e) Temperature dependence of resistivity for $ZrTe_2$. Inset: The evolution of the resistivity at low temperature. (f) Transversal Hall resistance of $ZrTe_2$ single crystal at 10 K.

At ambient pressure, $ZrTe_2$ adopts a hexagonal $CdI_2$-type structure with space group $P$-$3m$1 (No. 164) as shown in Fig. 1(a). The synthesized $ZrTe_2$ sample is characterized by the XRD experiments. The XRD pattern of $ZrTe_2$ single crystal is shown in Fig. 1(b). The (00l) plane is a natural cleavage facet of as-grown single crystals. The full width at half maximum (FWHM) of (002) peak is only 0.04° [inset of Fig. 1(b)], indicating the high quality of our samples. The $c$-axis lattice constant is 6.625 ± 0.005 Å, consistent with the previous reports[37]. The energy dispersive x-ray spectrometry (EDXS) data in Fig. S1 gives the ratio of Zr:Te as 1 : 2.01. Fig. 1(d) presents the band structure of $ZrTe_2$ calculated along high-symmetry lines in the first Brillouin zone (BZ) [Fig. 1(c)]. We can observe a band inversion around the $\Gamma$ point, confirming topological semimetal behaviors. The Fig. 1(e) exhibits the temperature dependence of resistivity for $ZrTe_2$ crystal. A metallic behavior is observed with decreasing temperature followed by the resistive upturn below ~ 6 K. the resistivity behavior shown here is in line with the previously reported data, which may derive from weak Kondo effect[37]. We further conduct the transversal Hall resistance at 10 K [Fig. 1(f)] and the ZrTe2 is dominated by electron-type carriers with the electron concentration $n_e \sim 2.97 \times 10^{21}$ cm$^{-3}$ at 10 K.

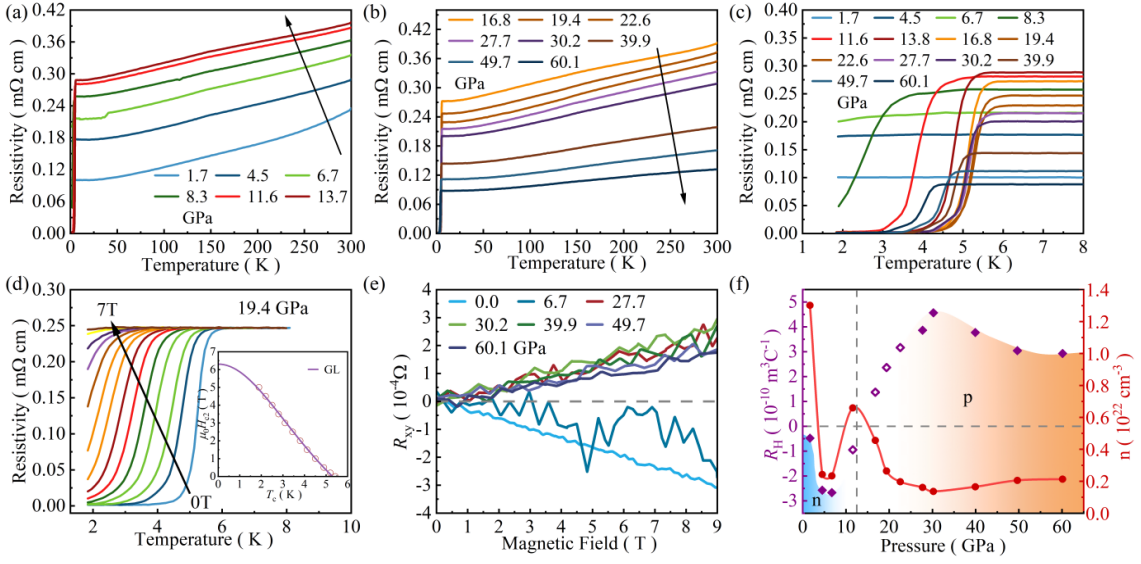

**FIG. 2**. Electrical resistivity of ZrTe2 as a function of temperature (a) below and (b) above 13.7 GPa. (c) Temperature-dependent resistivity of ZrTe2 in the vicinity of superconductivity. (d) Temperature dependence of resistivity under different magnetic fields for ZrTe2 at 19.4 GPa. The inset shows the results of Ginzburg-Landau fitting. (e) Hall resistance of ZrTe2 as a function of magnetic field under selected pressures at 10 K. (f) Pressure dependence of Hall coefficient and carrier concentration at 10 K for ZrTe2. Hollow points represent noise effects.

Under high pressure, we measure the temperature dependence of resistivity $\rho(T)$ for ZrTe2 crystals. As depicted in Fig. 2(a) and (b), ZrTe2 keeps metallic behavior up to 60.1 GPa, while the normal state of resistivity exhibits a non-monotonic evolution with increasing pressure. Increasing the pressure initially induces continuous enhancement of the overall magnitude of $\rho$ with a maximum occurring at 13.7 GPa. Upon further increasing the pressure, the resistivity starts to decrease gradually. As pressure increases up to 8.3 GPa, a sharp drop of resistivity in ZrTe2 is observed at the lowest temperature (experimental $T_{min}$ = 1.8 K), indicating the emergence of superconductivity, and zero resistivity is obtained when the pressure enhances to 11.6 GPa. The critical temperature $T_c$ reaches the maximum $T_c$ of 5.6 K around 19.4 GPa and then decreases with pressure, as plotted in Fig. 2(c) ($T_c$ is referred to $T_c^{onset}$ defined as the temperature at 90% of the residual resistivity in this paper). The measurements on different samples of ZrTe$_2$ from two independent runs provide reproducible and consistent results, confirming the superconductivity transition under pressure [Fig. S2]. To gain insights into the superconducting transition, we applied the magnetic field for ZrTe$_2$ subjected to 19.4 GPa. As shown in Fig. 2(d), $T_c$ is gradually suppressed with the enhancement of magnetic fields and the superconductivity extinguishes under the magnetic field $\mu_0H$ = 7 T. We tried to use the Ginzburge-Landau formula to fit the data [inset of Fig. 2(d)]. The estimation of $\mu_0H_{c2}$ at 0 K is ~ 6.1 T, and the Ginzburg-Landau coherence length $\xi_{GL}(0)$ is 7.3 nm. High-pressure Hall resistivity measurements were further carried out to extract the evolution of charge carriers in the pressurized ZrTe$_2$. Fig. 2(e) and Fig. S3 show the Hall resistivity curves $R_{xy}(H)$ measured at 10 K under various pressures. At low pressure region, the $R_{xy}(H)$ curve exhibits a negative slope, indicating an electron-dominated feature of the electrical transport. This is in agreement with the carriers type at ambient pressure. When the pressure is above 27.7 GPa, the Hall resistance slope becomes positive, suggesting the dominance of hole-type carriers. In the first run of Hall measurements, $R_{xy}$ is effected by the noise between 6.7 and 27.7 GPa due to the competition between two types of carriers caused by pressure gradient. We repeated the high-pressure Hall measurements with sodium chloride as pressure transmitting medium. Compared to previous Hall measurements, we reduced the noise influence, as shown in Fig. S3(b). The slope of $R_{xy}$ changes from negative to positive at 11.2 GPa, and the sign change of the $R_{xy}(H)$ demonstrates the change of charge carrier type, which indicates the change of the Fermi surface topology. This variation could be viewed as a signature of the Lifshitz transition[41, 45]. Pressure dependence of Hall coefficient and carrier concentration at 10 K are summarized in Fig. 2(f).

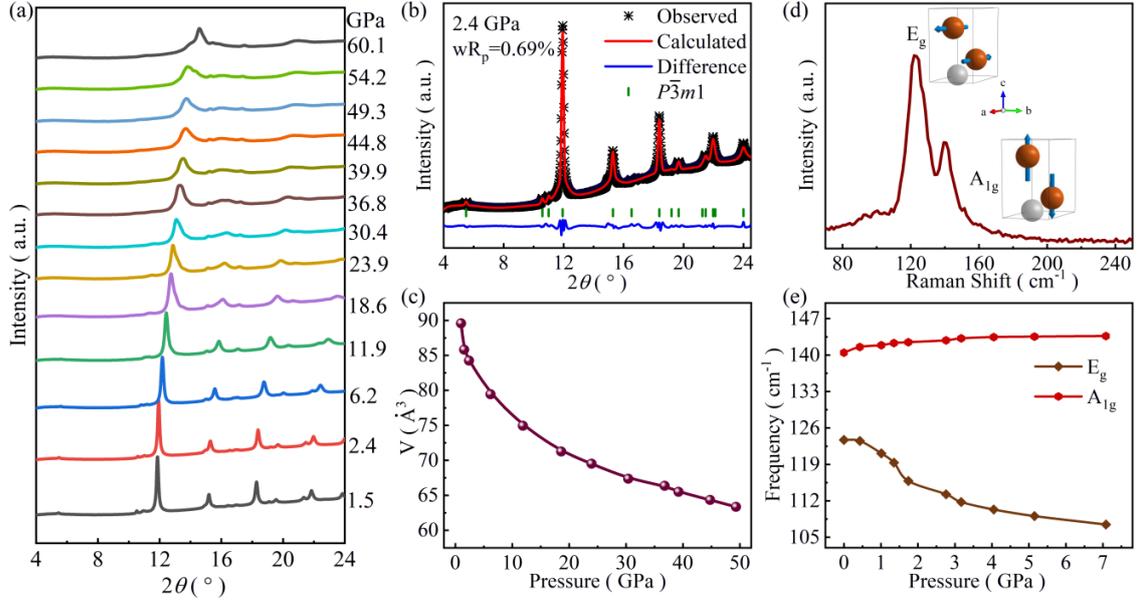

**FIG. 3**. (a) High pressure XRD patterns of ZrTe$_2$ up to 60.1 GPa at room temperature. The X-ray diffraction wavelength $\lambda$ is 0.6199 Å. (b) Rietveld refinement of XRD pattern at 2.4 GPa. The red solid line and black stars represent the calculated and experimental data, respectively, and the blue solid lines are the residual intensities. The vertical bars are the diffraction peak positions. (c) Pressure dependence of unit-cell volume. (d) Raman spectra of ZrTe$_2$ at ambient pressure (Zr: gray; Te: orange). (e) Pressure dependence of vibration modes frequencies of ZrTe$_2$.

To examine the thermodynamic stability of the ZrTe$_2$ phase and whether the pressure-induced SC is associated with structural phase transition, we performed *in situ* high-pressure powder XRD measurements at room temperature. Fig. 3(a) displays the high-pressure synchrotron XRD patterns of ZrTe$_2$ up to 60.1 GPa. A representative refinement at 2.4 GPa is presented in Fig. 3(b). All the diffraction peaks can be indexed well to ambient structure (space group *P*-3*m*1, No. 164) based on Rietveld refinement with General Structure Analysis System (GSAS) software package. All the XRD peaks continuously shift towards higher angles without new peaks appearing when the pressure increases up to 49.3 GPa, indicating the absence of structural phase transition in the pressurized ZrTe$_2$. Above 49.3 GPa, the signal intensity of the main peak deviates from the symmetry of *P*-3*m*1. We expect to study this structural transitions in the future. Fig. 3(c) shows the pressure (*P*) dependence of volume (*V*). Upon compression from 1.5 to 49.3 GPa, the overall volume decreases by 29% without volume collapse. In addition, We have performed single-crystal XRD under 5.8 and 14.3 GPa [Fig. S14 and Table S1]. The results of single crystal XRD demonstrate that ZrTe$_2$ retains $P$-$3m$1 up to 14.3 GPa. The stability of ZrTe$_2$ was also confirmed by *in situ* Raman spectroscopy measurements. As shown in Fig. 3(d), the Raman spectra at ambient pressure contain two characteristic peaks, which are due to the in-plane mode $E_g$ and the out-of-plane mode $A_{1g}$ of the ZrTe$_2$ structure; this is also in agreement with a previous report[46]. The frequencies of both vibrational modes move gradually without discontinuities as pressure increases [Fig. S4] indicating the robustness of structure in the whole studied pressure range at room temperature. Interestingly, $E_g$ mode displays the opposite trend and shows redshift behavior when the pressure is raised. This pressure-induced phonon softening is probably associated with emergence of superconductivity in the pressurized ZrTe$_2$.

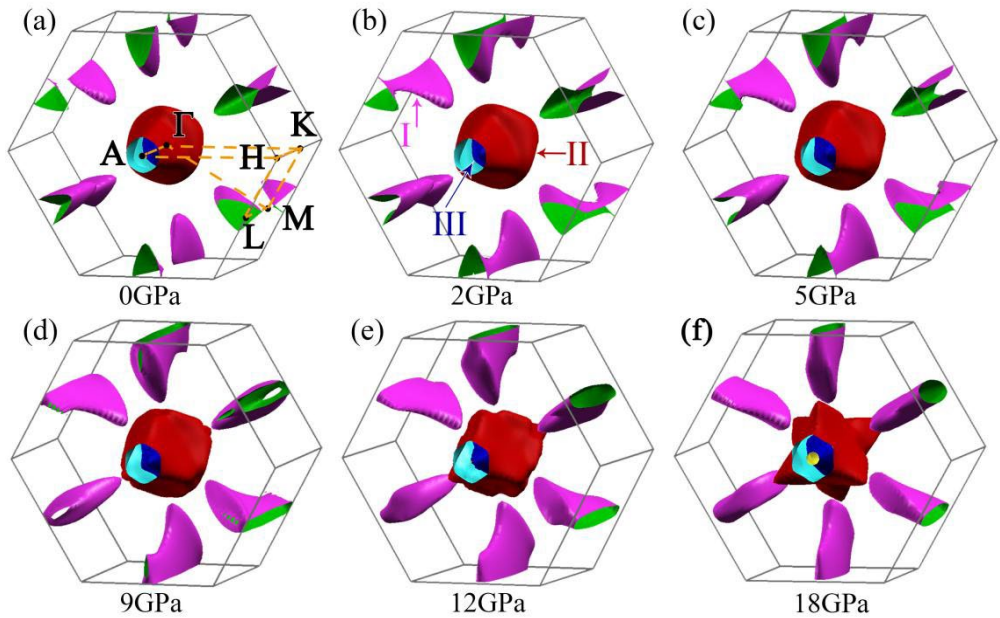

**FIG. 4**. The Fermi surface of ZrTe$_2$ at various pressures. The Brillouin zone path is shown in (a), and different Fermi surfaces are indexed as I, II and III in (b), respectively.

To further understand the transport behavior of ZrTe$_2$ under high pressure, we first conducted the Fermi surface calculations at various pressures [Fig. 4]. The Fermi surfaces are indexed as I, II and III in Fig. 4(b). The decomposed Fermi surfaces are plotted in Fig. S5-S8, and the Fermi surface IV is enclosed by the Fermi surface III. The Fermi surface I forms a connection around 2 GPa [Fig. 4(b)]. The pocket enlarges with the pressure and closes around 12 GPa [Fig. 4(e)]. The Fermi surface II transforms from shuttle shape to David-star shape at 18 GPa [Fig. S6]. The hexagon [Fig. S7(c)] in Fermi surface III becomes a David-star at 12 GPa [Fig. S7(e)], and an opening emerges at *A* point under 18 GPa [Fig. S7(f)]. The Fermi surface IV emerges around 5 GPa [Fig. S5(c)] and vanishes about 12 GPa [Fig. S5(e)]. The evolution of the Fermi surface under high pressure could be the signature of the Lifshitz transitions, and the reshape of Fermi surface I, III and IV at 12 GPa is in line with the normal state anomalies in our resistivity measurements.

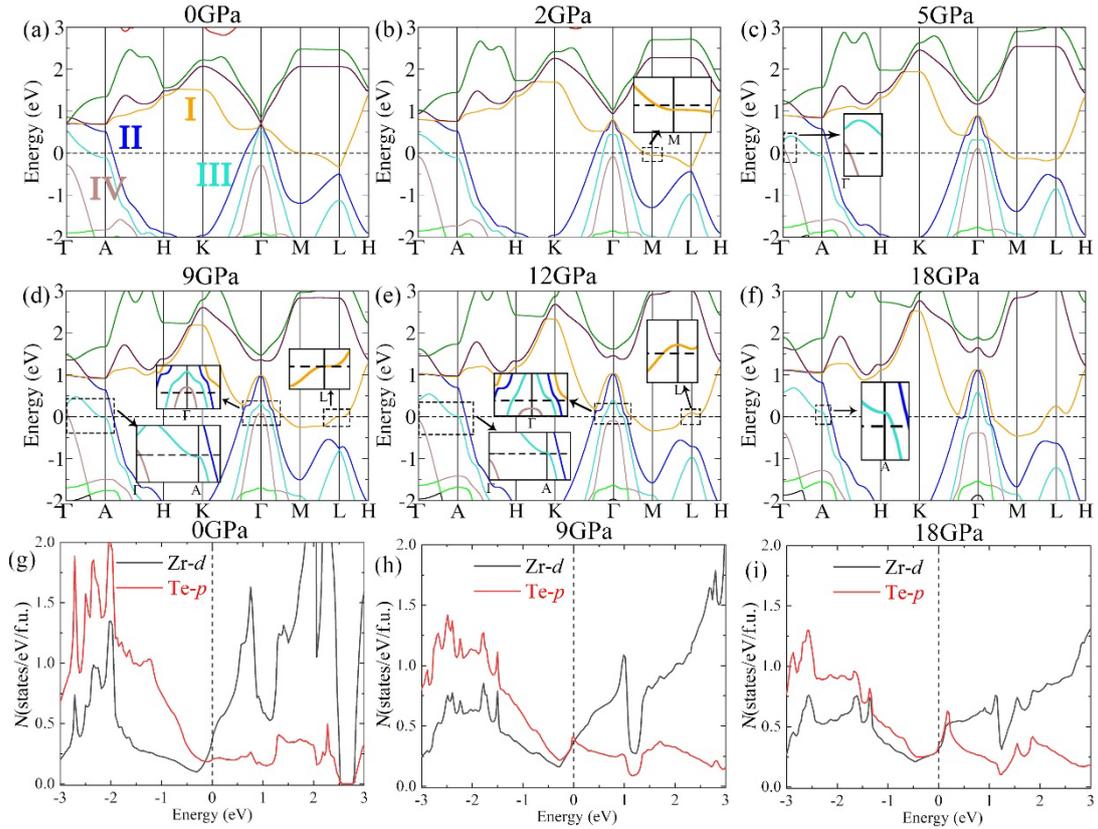

**FIG. 5**. (a)-(f) The band structures of ZrTe$_2$ at various pressures. The orange, blue, cyan and brown bands are indexed as band I, II, III and IV, respectively. The insets show the details of the band structures. (g)-(i) The partial density of states (PDOS) at various pressures. The black and red solid lines are the PDOS of the *d* electrons of Zr atoms and the *p* electrons of Te atoms, respectively.

Next, we calculated the band structures of ZrTe$_2$ at various pressures and the details are shown in Figs. 5(a)-(f). The band indexes I, II, III, and IV are in line with the Fermi surfaces. The band I crosses the Fermi energy along *Γ-M* at 2 GPa [Fig. 5(b)], causing the connection in Fig. 4(b), and the pocket enclosing at *L* point [Fig. 4(e)] can be identified to a *p*-type conversion of band I [Fig. 2(e)-(f)]. As for the David-star in the Fermi surface II, the band structures on the $k_z = 0.25$ plane [Figs. S6(g)-(i)] indicates a saddle point around the Fermi energy under 18 GPa [Fig. S6(i)]. The reshape of the Fermi surface III [Fig. S7] is owing to the saddle point along $\Gamma$-$M$ [Fig. 5(e)], and the opening at 18 GPa is due to the $p$-type crossing at the Fermi energy around $A$ point [Figs. 5(e)-(f)]. The emerging and vanishing of Fermi surface IV originates from the variation of band IV around $\Gamma$ point from 5 GPa to 12 GPa [Fig. 5(c)-(e)]. Thus, band structures calculation provides details for the Lifshitz transitions of Fermi surface topology, and the $p$-type conversion at $A$ [Fig. S10] and $L$ [Fig. S9] points is in line with the charge carrier type transition under high pressure.

Moreover, we calculated the orbital contribution around $A$, $\Gamma$ and $L$ points [Fig. S9-S11]. Around the Fermi energy, the main distribution is from the $p$ electrons of Te atoms, such as the band I at $L$ point [Fig. S9], band III at $A$ point [Fig. S10] and band IV at $\Gamma$ point [Fig. S11]. Accordingly, we calculated the partial density of states (PDOS) at various pressures [Figs. 5(g)-(h)]. The PDOS is more diverged under high pressure, and a peak emerges on the Fermi energy around 9 GPa. This peak is from the Te-$p$ electrons and could contribute to the pressure induced superconductivity. Besides, we calculated the charge density between the inter-layer and intra-layer Te atoms under high pressure [Fig. S12]. Compared with intra-layer Te atoms, more electrons are distributed between inter-layer Te atoms. This is empirically true since the inter-layer distance is easier to compress in layered TMDCs. Such novel bonds explain the DOS peak of Te-$p$ electrons around the Fermi energy, which could be favorable for the superconducting transition through electron-phonon coupling. Hence, our results demonstrated that the anisotropic compression behaviors in ZrTe$_2$ causes the redistribution of Te-$p$ electrons. It leads to the reshape of the band structures and the Fermi surface topology, which is in agreement with the transport anomalies of normal state and the carriers type conversion under high pressure. The anisotropic compression causes the bonding and PDOS elevation around Fermi energy of Te-$p$ electrons as well, which is in consistent with the pressure induced superconductivity in our experiments.

TABLE I. The $\mathbb{Z}_2$ invariant of band II under different pressures.

| Pressure (GPa) | Time reversal invariant planes | | | | | | $\mathbb{Z}_2$ index $(\nu_0;\nu_1\nu_2\nu_3)$ |
|---|---|---|---|---|---|---|---|
| | $k_x=0.0$ | $k_y=0.5$ | $k_y=0.0$ | $k_y=0.5$ | $k_z=0.0$ | $k_z=0.5$ | |
| 0 | 0.0 | 0.0 | 0.0 | 0.0 | 1.0 | 0.0 | (1;000) |
| 2 | 1.0 | 0.0 | 1.0 | 0.0 | 1.0 | 1.0 | (0;001) |
| 5 | 0.0 | 0.0 | 0.0 | 0.0 | 0.0 | 1.0 | (1;001) |

| | | | | | | |
|---|---|---|---|---|---|---|
| 7  | 1.0 | 0.0 | 1.0 | 0.0 | 0.0 | 1.0 | (1;001) |
| 9  | 1.0 | 0.0 | 1.0 | 0.0 | 0.0 | 1.0 | (1;001) |
| 12 | 1.0 | 0.0 | 1.0 | 0.0 | 0.0 | 1.0 | (1;001) |
| 18 | 0.0 | 0.0 | 0.0 | 0.0 | 0.0 | 1.0 | (1;001) |
| 24 | 0.0 | 0.0 | 0.0 | 0.0 | 0.0 | 1.0 | (1;001) |
| 30 | 0.0 | 0.0 | 0.0 | 0.0 | 0.0 | 1.0 | (1;001) |
| 50 | 0.0 | 0.0 | 0.0 | 0.0 | 0.0 | 0.0 | (0;000) |

Meanwhile, we observed a gap opening at $\Gamma$ point [Fig. S11] and the band inversion at $L$ point [Fig. S9], suggesting potential topological properties. We calculated the $\mathbb{Z}_2$ invariant of band I and II up to 50 GPa. The detailed results are shown in Table I. The band I keeps topologically trivial within the pressure range. For band II, it transforms from topologically non-trivial state at 30 GPa to trivial state at 50 GPa, and the variation of $\mathbb{Z}_2$ invariant around 2 GPa (1→0→1) suggests the topological states transition. This is similar to the results observed in $\beta$-Bi$_4$I$_4$ [47]. The surface states on (001) plane at various pressures are shown in Fig. 6 and Fig. S13. We could observe the split and cross of surface states around $\bar{M}$ point with the pressure increasing, while, the surface states around the $\bar{\Gamma}$ point are more complex as shown in Fig. S9. Therefore, the topological properties of ZrTe$_2$ could be modulated by high pressure. More importantly, our results shown here demonstrate the coexistence of non-trivial topology and superconductivity in ZrTe$_2$ upon compression. Our study will stimulate further studies, such as the quantum oscillations[48, 49] and Josephson effect[50, 51] under high pressure, to explore potential topological superconductivity and Majorana fermions.

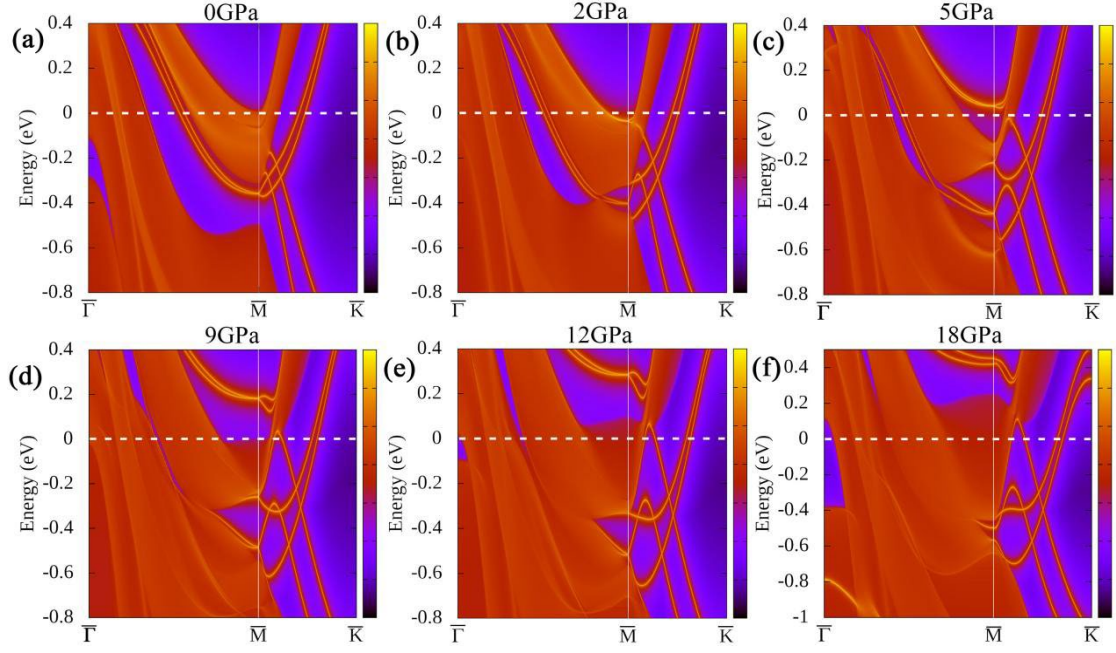

**FIG. 6**. The surface states of ZrTe$_2$ on the (001) plane around $\bar{M}$ point at various pressures.

Based on the above resistivity, XRD, and theoretical calculation, the *T-P* phase diagram is summarized in Fig. 7. These results demonstrate that high pressure dramatically alters both topological and transport properties of ZrTe$_2$. Its crystal sustains a hexagonal CdI$_2$-type structure under high pressures up to 60.1 GPa, while applied pressure induces multiple topological quantum phase transitions in ZrTe$_2$. More interestingly, superconductivity emerges around 8.3 GPa and $T_c$ reaches maximum of 5.6 K around 19.4 GPa, showing a typical dome-like evolution. The combined theoretical calculations and *in situ* high pressure measurements demonstrate the topologically non-trivial state is accompanied by the appearance of superconductivity, making ZrTe$_2$ possible platform to study topological superconductivity.

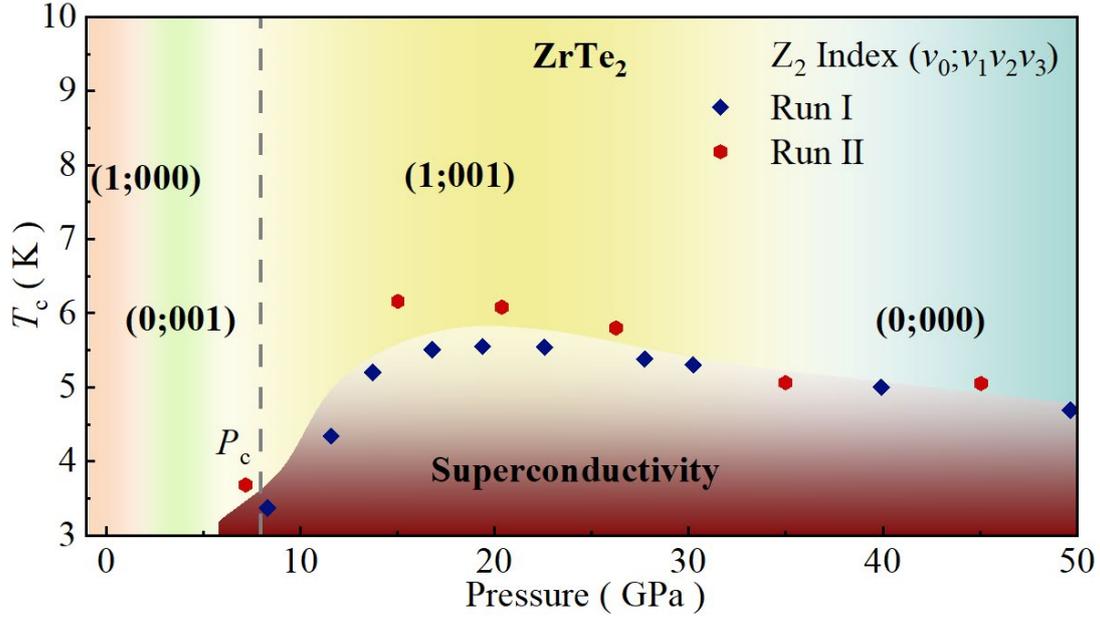

**FIG. 7**. Phase diagram of ZrTe$_2$. Superconducting transition emerges at around 8.3 GPa and maintains up to 60 GPa. Topological state transitions are shown as the variation of $\mathbb{Z}_2$ index. A topological states transition arises around 2 GPa and topologically non-trivial state coexists with superconducting state up to 30 GPa.

## 3. Conclusion

In summary, we discovered pressure-induced superconductivity in topological TMDC ZrTe$_2$ by combining experimental and theoretical investigations. High pressure dramatically alters the electronic state, and a pressure-induced Lifshitz transition is evidenced by the change of charge carrier type as well as the Fermi surface. Superconductivity is observed in ZrTe$_2$ at large pressure region with a dome-shape evolution. Theoretical calculations indicated that ZrTe$_2$ experiences multiple pressure-induced topological quantum phase transitions, which coexists with superconductivity. Our results demonstrate that ZrTe$_2$ with a nontrivial topology of electronic states display new ground states upon compression and have potential applications in next-generation spintronic devices.

Supporting Information

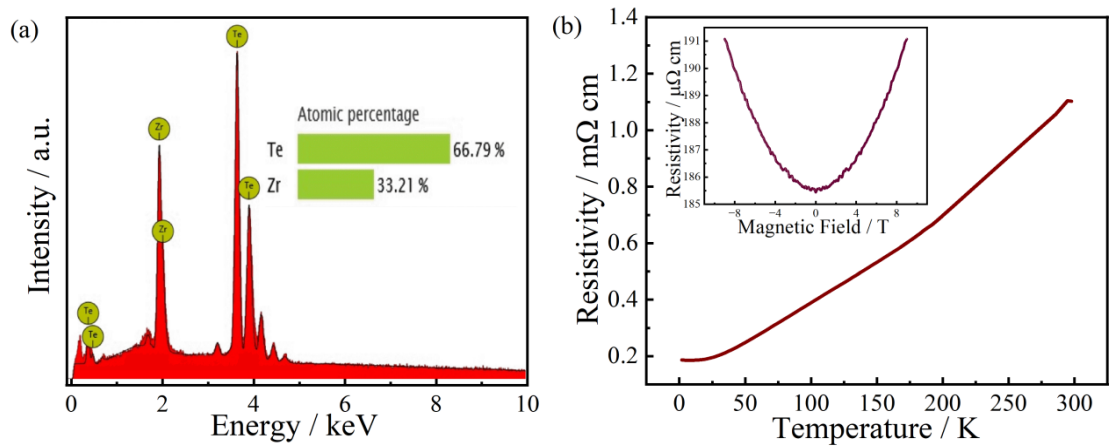

**FIG. S1**. (a) Energy-dispersive x-ray spectroscopy of ZrTe$_2$ (b) Temperature dependent resistivity of ZrTe$_2$ single crystal. Inset: magnetoresistivity of ZrTe$_2$ when ab-plane is perpendicular to magnetic field.

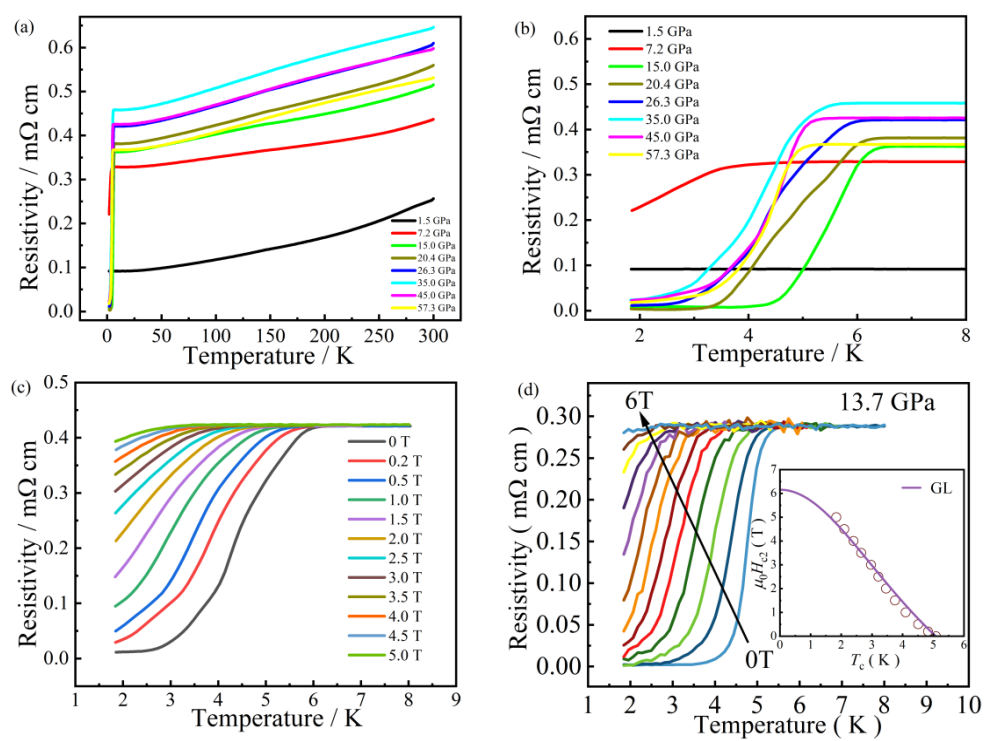

**FIG. S2**. (a) Electrical resistivity of ZrTe$_2$ as a function of temperature under pressures in run II. (b) Temperature-dependent resistivity of ZrTe$_2$ in the vicinity of the superconducting transition in run II. (c) Temperature dependence of resistivity under different magnetic fields for ZrTe$_2$ at 26.3 GPa in run II. (d) Temperature dependence of resistivity under different magnetic fields for ZrTe$_2$ at 13.7 GPa in run II.

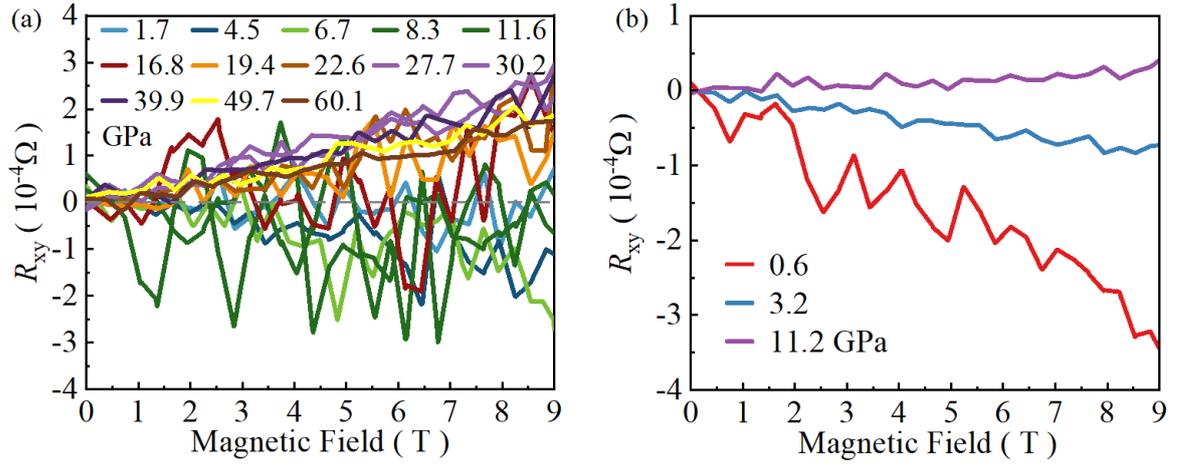

**FIG. S3.** (a) Hall resistance of ZrTe$_2$ as a function of magnetic field under various pressures at 10 K. (b) Hall resistance of ZrTe$_2$ as a function of magnetic field up to 11.2 GPa with sodium chloride as pressure transmitting medium at 10 K.

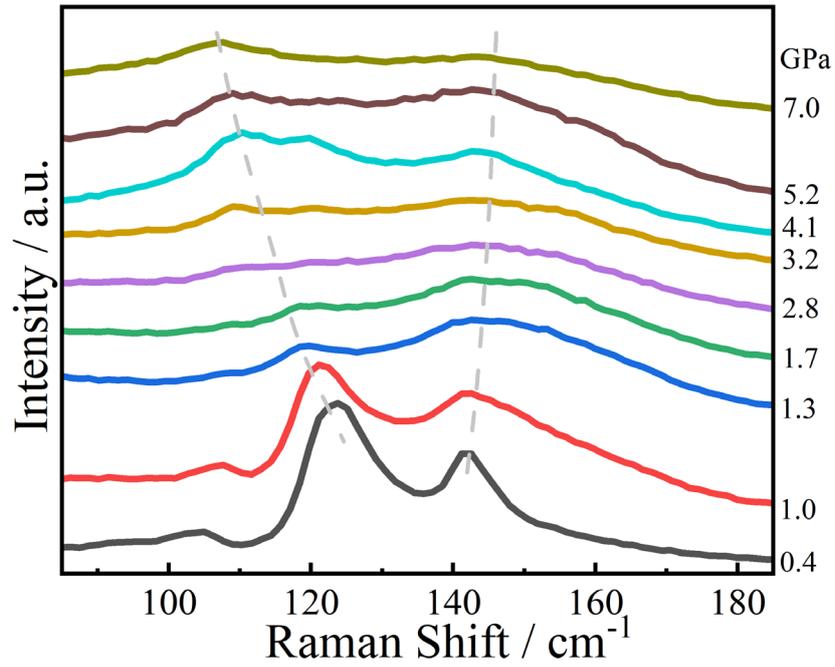

**FIG. S4.** Raman spectra at various pressures for ZrTe$_2$ at room temperature.

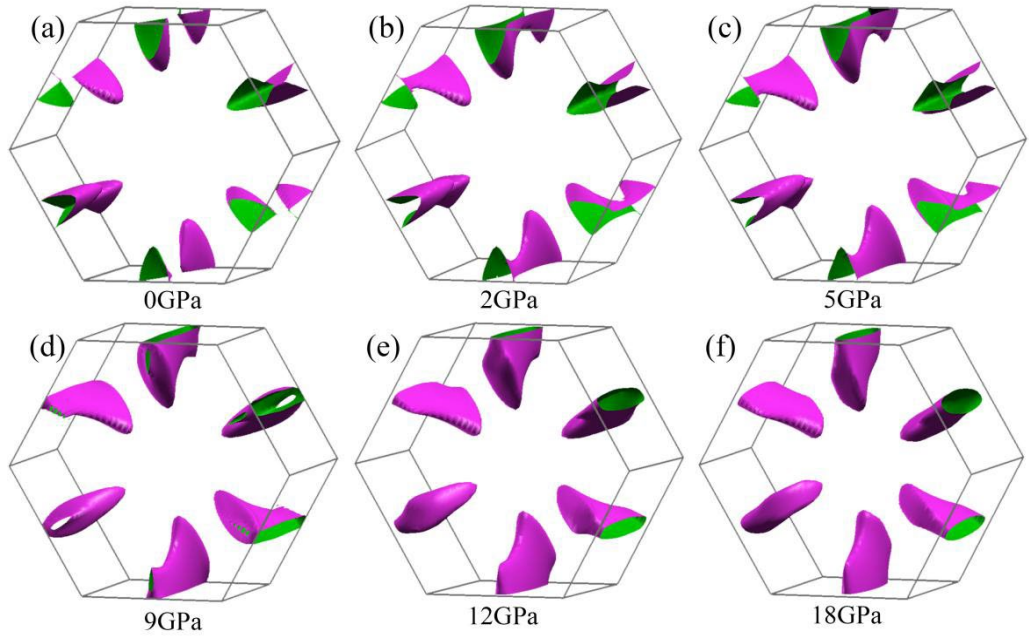

**FIG. S5**. The Fermi surface of the band I under different pressures.

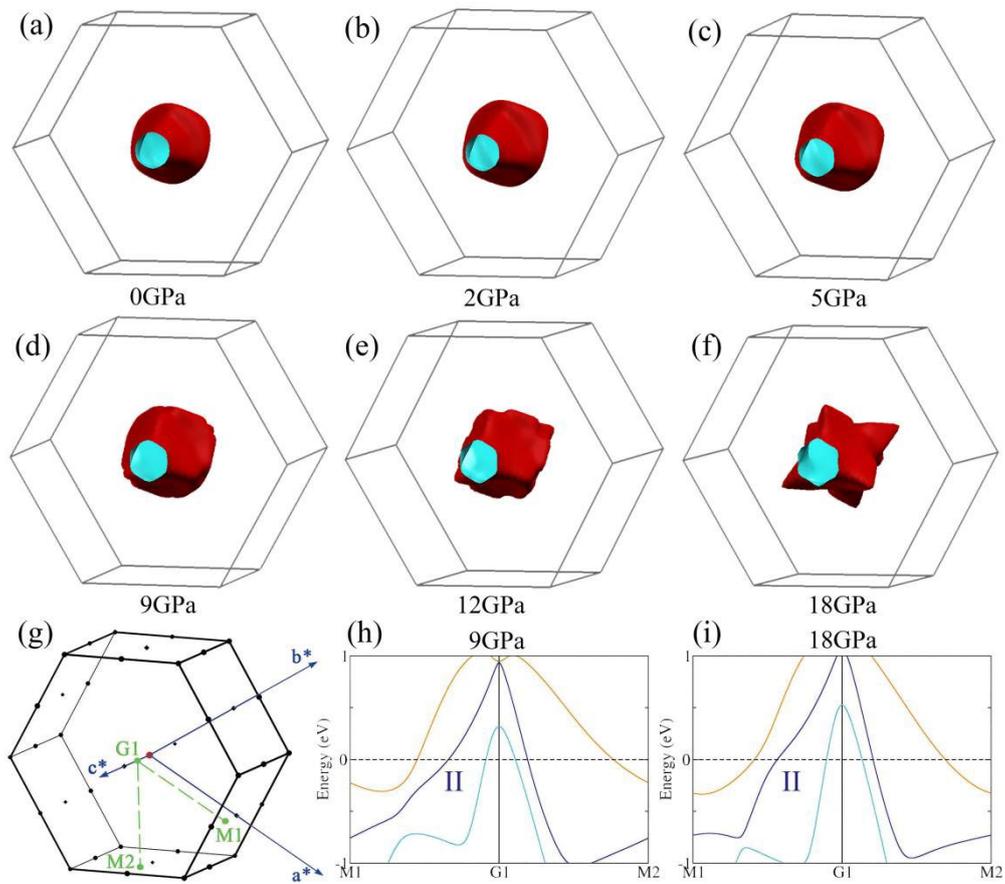

**FIG. S6**. (a)-(f) The Fermi surface of the band II under different pressures. (g) The Brillouin zone path on the $k_z = 0.25$ plane. The band structures at 9 GPa (h) and 18 GPa (i), the arrow points to a saddle point around the Fermi level at 18GPa (i).

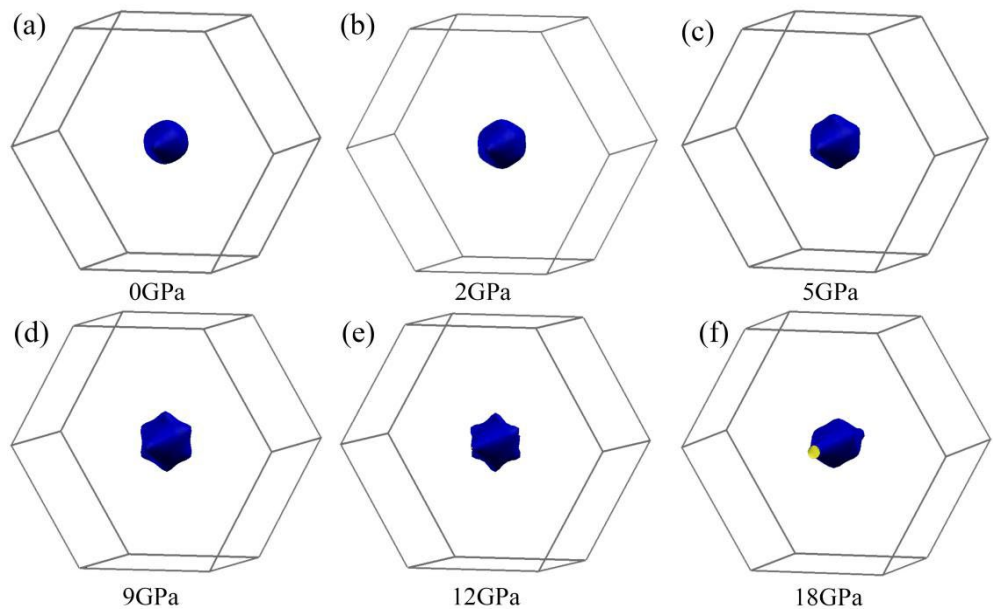

**FIG. S7**. The Fermi surface of the band III under different pressures.

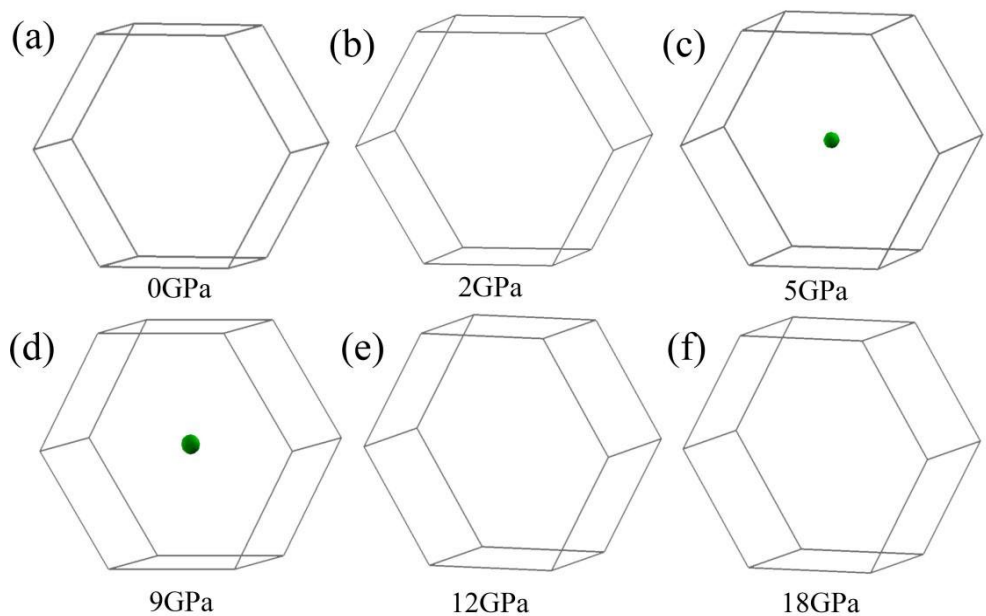

**FIG. S8**. The Fermi surface of the band IV under different pressures.

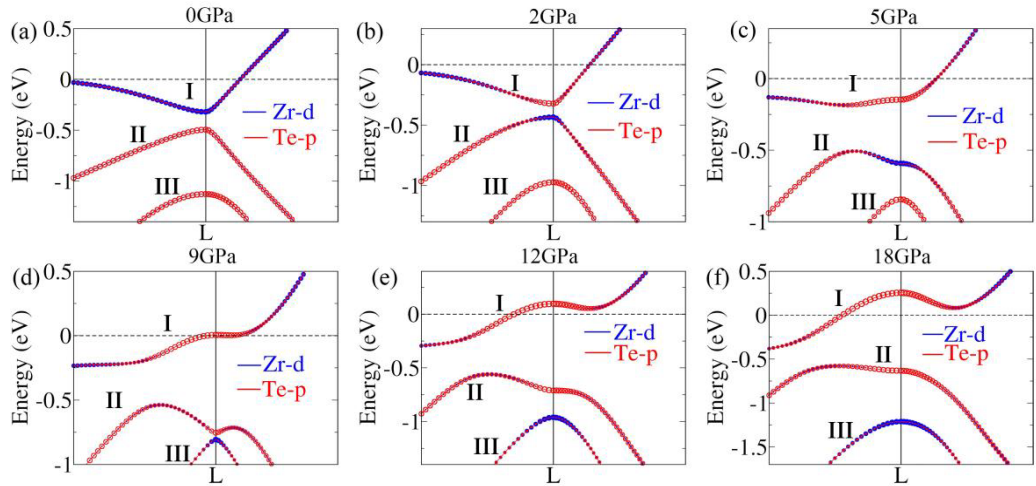

**FIG. S9**. The details of the band structures at *L* point, the blue points and the red points are the contribution from the *d* electrons of Zr atoms and the *p* electrons of Te atoms.

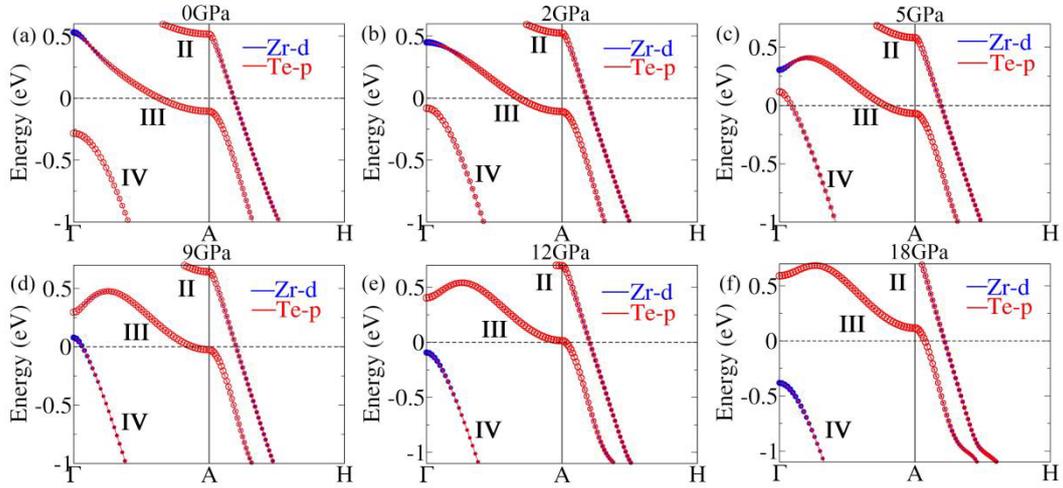

**FIG. S10**. The details of the band structures at *A* point, the blue points and the red points are the contribution from the *d* electrons of Zr atoms and the *p* electrons of Te atoms.

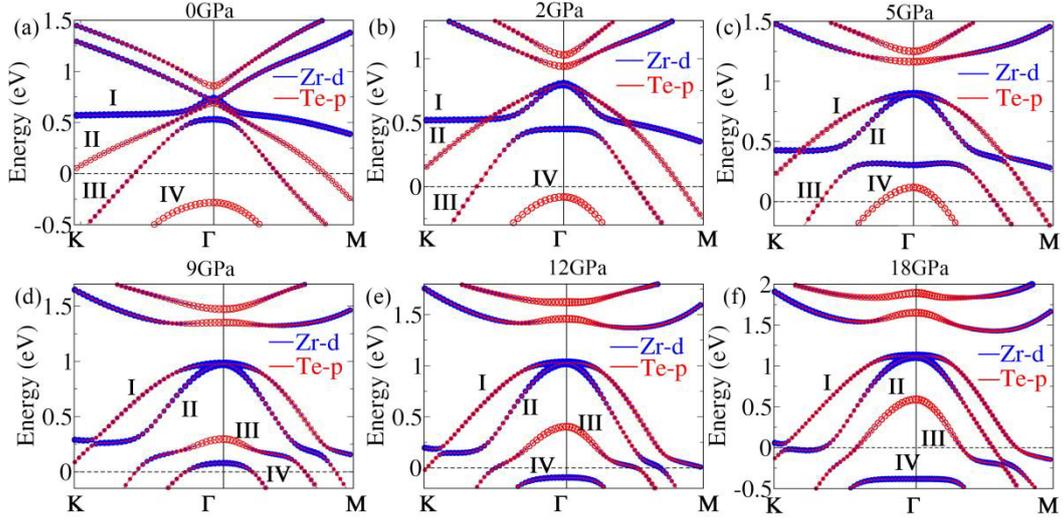

**FIG. S11**. The details of the band structures at $\Gamma$ point, the blue points and the red points are the contribution from the $d$ electrons of Zr atoms and the $p$ electrons of Te atoms.

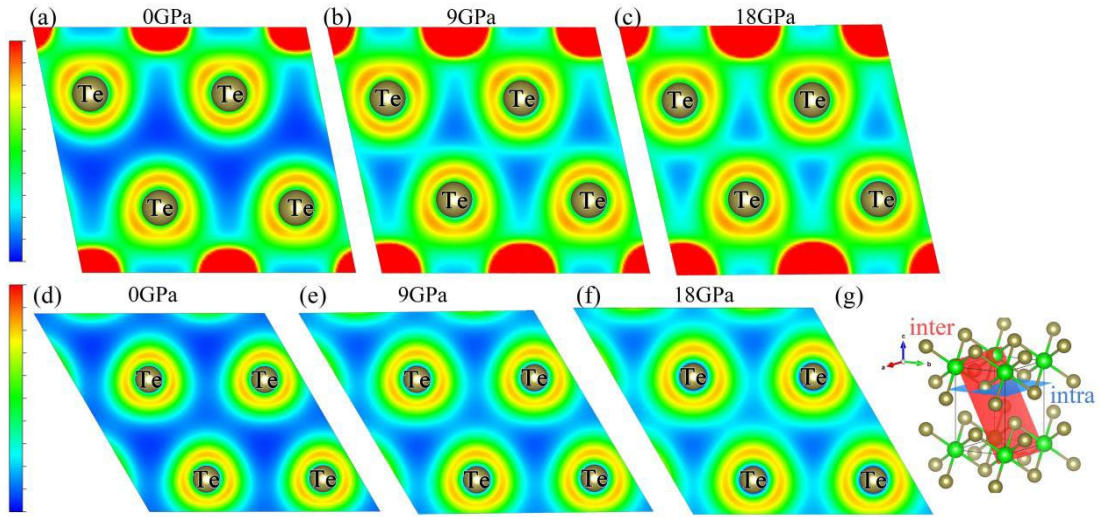

**FIG. S12**. The charge density between the inter-layer Te atoms (a)-(c) and the intra-layer Te atoms (d)-(f), the corresponding planes are in (g).

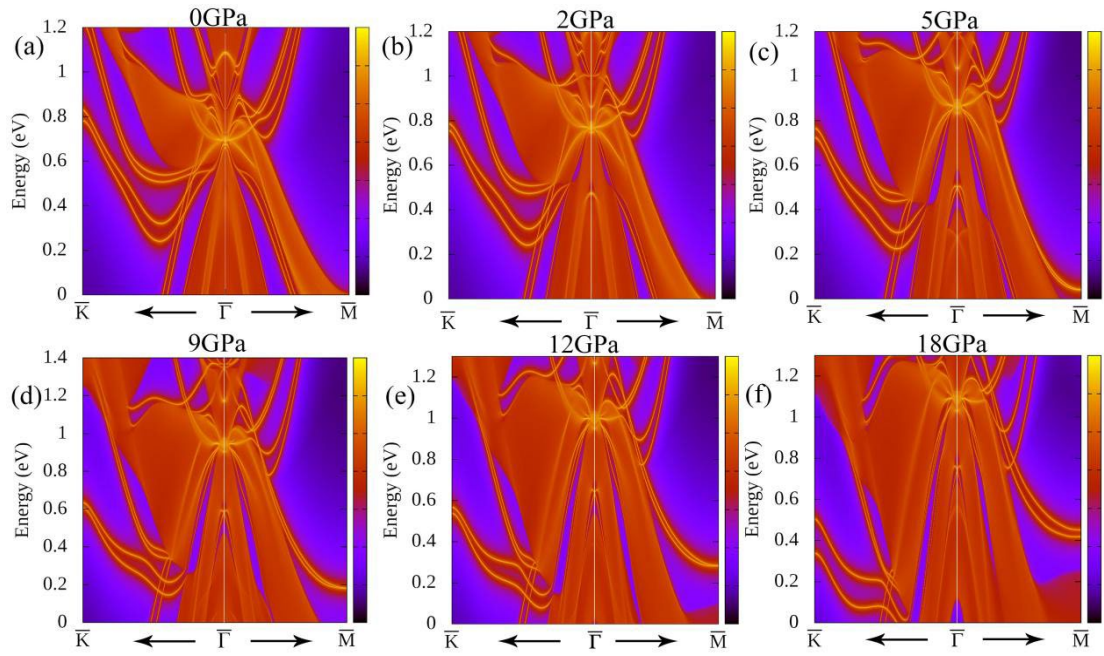

**FIG. S13**. The surface states on the (001) plane around $\bar{\Gamma}$ point at various pressures.

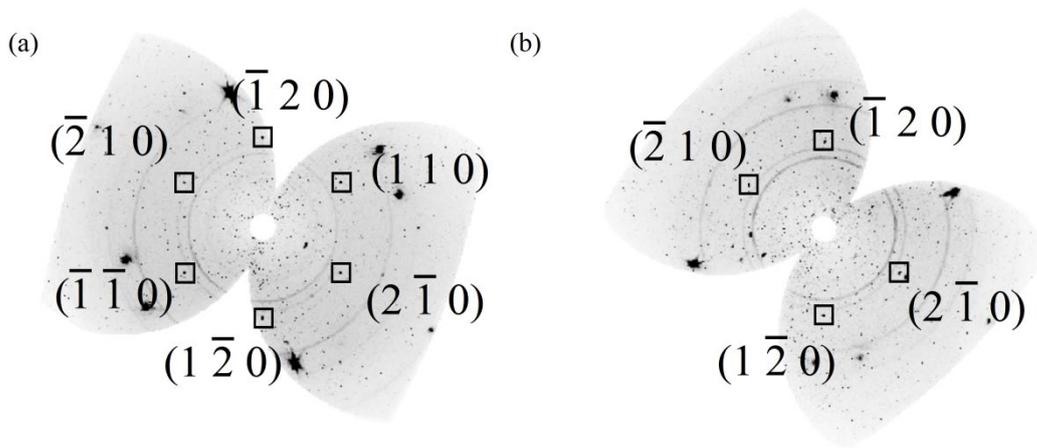

**FIG. S14**. single crystal XRD pattern of ZrTe$_2$ under (a) 5.8 GPa and (b) 14.3 GPa.

Table R1. Sample data and structure refinement for ZrTe$_2$ under 5.8 and 14.3 GPa.

| Pressure of ZrTe$_2$ | 5.8 GPa | 14.3 GPa |
|---|---|---|
| Formula weight | 364.42 | 364.42 |
| Temperature | 296(2) K | 296(2) K |
| Wavelength | 0.71073 Å | 0.71073 Å |
| Crystal system | Trigonal | Trigonal |
| Space group | $P\bar{3}m1$ | $P\bar{3}m1$ |
| Unit cell dimensions | a = 3.9553 (11) Å, α = 90°<br>b = 3.9553 (11) Å, β = 90°<br>c = 6.636 (7) Å, γ = 120° | a = 3.926(2) Å, α = 90°<br>b = 3.926(2) Å, β = 90°<br>c = 6.552(2) Å, γ = 120° |
| Volume | 89.91 (11) Å$^3$ | 87.46(9) Å$^3$ |
| Z | 1 | 1 |
| Density (calculated) | 6.398 Mg/m$^3$ | 6.578 Mg/m$^3$ |
| Absorption coefficient | 18.67 mm$^{-1}$ | 19.199 mm$^{-1}$ |
| F(000) | 144 | 144 |
| Crystal size | 0.035 x 0.032 x 0.027 mm$^3$ | 0.035 x 0.032 x 0.027 mm$^3$ |
| θ range for data collection | 6.0 to 22.2° | 6.0 to 22.38° |
| Index ranges | -4<=h<=4, -3<=k<=3, -1<=l<=1 | -4<=h<=4, -4<=k<=4, -1<=l<=1 |
| Reflections collected | 270 | 156 |
| Independent reflections | 18 [R(int) = 0.0570] | 18 [R(int) = 0.1413] |
| Coverage of independent reflections | 29.0 % | 30.0 % |
| Refinement method | Full-matrix least-squares on F2 | Full-matrix least-squares on F2 |
| Data / restraints / parameters | 18 / 6 / 6 | 18 / 12 / 6 |
| Goodness-of-fit | 1.588 | 1.020 |
| Final R indices [>2σ(I)] | R1 = 0.0288, wR2 = 0.0559 | R1 = 0.0781, wR2 = 0.1630 |
| R indices [all data] | R1 = 0.0451, wR2 = 0.0608 | R1 = 0.1196, wR2 = 0.1980 |
| Largest diff. peak and hole | 1.044 and -0.623 e.Å$^{-3}$ | 1.401 and -1.364 e.Å$^{-3}$ |